\documentclass[aps,prl,twocolumn,final,letterpaper]{revtex4}

\usepackage{graphicx}   
\usepackage{import}                         
\usepackage{epstopdf}
\usepackage{amsmath} 
\usepackage{bm}
\usepackage{amssymb}
\usepackage{quotes}
\usepackage{transparent}
\usepackage{dcolumn}
\usepackage[usenames, dvipsnames]{color}
\usepackage{multirow}
\usepackage{cancel} 
\usepackage{mdframed}
\usepackage{color}
\usepackage{bm}
\usepackage{setspace}
\usepackage{dsfont}
\usepackage{braket}
\usepackage{slashed}
\usepackage{float}
\usepackage{soul, color} 
\soulregister\ref{7}  
\soulregister\cite{7} 
\renewcommand{\st}[1]{}

\newcommand{\rv}{\mathbf{r}}

\newcommand{\kv}{\mathbf{k}}

\newcommand{\appropto}{\mathrel{\vcenter{
  \offinterlineskip\halign{\hfil$##$\cr
    \propto\cr\noalign{\kern.2pt}\sim\cr\noalign{\kern-2.5pt}}}}}

\newcommand{\Pv}{\mathbf{P}}

\newcommand{\Ev}{\mathbf{E}}

\begin{document}
\rmfamily

\title{Two photon emission from superluminal and accelerating index perturbations}
\author{Jamison Sloan$^{1}$, Nicholas Rivera$^{2}$, John D. Joannopoulos$^{2}$, and Marin Solja\v{c}i\'{c}$^{2}$}

\affiliation{$^{1}$ Department of Electrical Engineering and Computer Science, Massachusetts Institute of Technology, Cambridge, MA 02139, USA \\
$^{2}$ Department of Physics, Massachusetts Institute of Technology, Cambridge, MA 02139, USA}

\noindent	

\begin{abstract}
Sources of photons with controllable quantum properties such as entanglement and squeezing are desired for applications in quantum information, metrology, and sensing. However, fine-grained control over these properties is hard to achieve, especially for two-photon sources. Here, we propose a new mechanism for generating entangled and squeezed photon pairs using superluminal and/or accelerating modulations of the refractive index in a medium. By leveraging time-changing dielectric media, where quantum vacuum fluctuations of the electromagnetic field can be converted into photon pairs, we show that energy- and momentum-conservation in multi-mode systems give rise to frequency and angle correlations of photon pairs which are controlled by the trajectory of the index modulation. These radiation effects are two-photon analogues of Cherenkov and synchrotron radiation by moving charged particles such as free electrons. We find the particularly intriguing result that synchrotron-like radiation into photon pairs exhibits frequency correlations which can enable the realization of a heralded single photon frequency comb. We conclude with a general discussion of experimental viability, showing how solitons, ultrashort pulses, and nonlinear waveguides may enable pathways to realize this two-photon emission mechanism. For completeness, we discuss in the Supplementary Information how these effects, sensitive to the local density of photonic states, can be strongly enhanced using photonic nanostructures. As an example, we show that index modulations propagating near the surface of graphene produce entangled pairs of graphene plasmons with high efficiency, leading to additional experimental opportunities.
\end{abstract}

\maketitle

\section{Introduction}
New methods for the controlled generation of entangled photon pairs and heralded single photons \cite{barz2010heralded} are of high interest for applications in quantum optics \cite{loudon1987squeezed}, quantum information \cite{o2007optical}, communications \cite{gisin2007quantum}, and sensing \cite{shapiro2008computational, degen2017quantum, d2003quantum}. Broadly, these sources rely on processes such as two-photon spontaneous emission from atoms \cite{shapiro1959metastability, breit1940metastability, rivera2017making}, semiconductors \cite{hayat2008observation, hayat2011applications, nevet2010plasmonic}, quantum dots \cite{ota2011spontaneous}, free electrons \cite{frank1960optics, rivera2019light}, and parametric down conversion in nonlinear crystals \cite{boyd2019nonlinear}. There has also been interest in $n$-photon emitters to create states of light that have potential applications in quantum information processing and medicine \cite{munoz2014emitters, bin2020n}. Any process involving more than one photon typically occurs with low efficiency, and can be difficult to control, particularly when the emission is into a system with many modes. As a result, new concepts for controllable sources of entangled photons remain in high demand.

Since entangled photon pairs are a fundamentally non-classical state of light, they are necessarily created through quantum processes involving matter. One particularly interesting case of this comes from time-varying optical media, whose dielectric properties are actively modulated in time \cite{law1994effective, lustig2018topological, zurita2009reflection,chu1972wave,harfoush1991scattering,fante1971transmission,holberg1966parametric}, giving rise to a wide array of classical electromagnetic effects which are the subject of many current investigations. In the quantum realm, time-varying photonic systems can spontaneously excite photons from the vacuum. This occurs because when a quantum system is varied in time, the original ground state of the system (containing no photons) may evolve into a quantum state which does contain photons. Phenomena which have been described this way include the dynamical Casimir effect \cite{moore1970quantum, dodonov2010current, wilson2011observation}, spontaneous parametric down-conversion (SPDC) in nonlinear materials \cite{louisell1961quantum, boyd2019nonlinear}, photon emission from rotating bodies \cite{maghrebi2012spontaneous}, the Unruh effect for relativistically accelerating bodies  \cite{yablonovitch1989accelerating, crispino2008unruh, fulling1976radiation, unruh1984happens}, Hawking radiation from black holes \cite{hawking1975particle, unruh1976notes}, and even particle production in the early universe \cite{shtanov1995universe}. These phenomena are linked together by the common thread of parametric amplification of vacuum fluctuations \cite{nation2012colloquium}.

In this work, we introduce a new concept for generating entangled photon pairs based on two-photon spontaneous emission from superluminal and accelerating permittivity perturbations in a medium. As some spatially localized index perturbation $\Delta\varepsilon$ is sent on some trajectory through a medium, electromagnetic vacuum fluctuations of the background medium induce spontaneous emission of entangled photon pairs. Depending on the background structure, light takes the form of generalized photons in a medium (e.g. free photons, photons in a bulk, cavity modes, photonic crystal modes, surface polaritons, etc.). We find that these fast index perturbations generate radiation somewhat similar to that created by a charged particle on the same trajectory. This correspondence allows us to characterize our two-photon processes as quantum analogs of free electron processes such as Cherenkov and synchrotron radiation (Fig. 1). 

Although these two-photon processes bear some similarities to their free electron analogs (due to energy and momentum conservation), key differences emerge in the spectrum and statistics of the emitted radiation. While an electron undergoing Cherenkov radiation emits photons into a ubiquitous cone, our system can emit photons across a broad angular spectrum, including backwards. We also show that an index perturbation moving in a circular trajectory emits ``synchrotron-like'' radiation, but with frequency comb-like entanglement between photon pairs. This concept alludes to the possibility of an entangled pair source where one photon is measured, heralding the other photon into a state which is a superposition over harmonics of a frequency comb. In the Supplementary Information (S.I.), we extend these concepts to nanostructured media, which can greatly increase the efficiency of these processes. As an example we demonstrate the Cherenkov concept in a system where a superluminal index perturbation propagates near the surface of graphene, which is of current interest due to its surface plasmon modes which exhibit high confinement, and can be tuned by electrical gating. We show that the plasmonic Cherenkov effect is more efficient than that in a uniform medium by 4 orders of magnitude.

In this new mechanism of radiation, the trajectory of index perturbations directly influences the frequency and spectral correlations exhibited between photons within each entangled pair. Thus our work could eventually lead to new controllable sources of entangled photons, which might be integrated into nanophotonic platforms such as polaritonic surfaces or ring resonators. Our concepts may also be particularly relevant in the wake of current interest in ``spatiotemporal metamaterials'' which control the flow of light using time as a new degree of freedom \cite{engheta2021metamaterials, caloz2019spacetime}. At microwave frequencies, spatiotemporal metamaterials built from time-modulated superconducting qubits could enable generation of entangled mirowave photons using these effects. At optical frequencies, index perturbations on various pre-defined trajectories can be realized by pulses propagating in nonlinear waveguides, fibers, or metasurfaces. In the section titled ``Experimental Outlook,'' we provide a detailed discussion of potential opportunities for realizing these mechanisms experimentally.

\begin{figure}[t]
    \centering
    \includegraphics{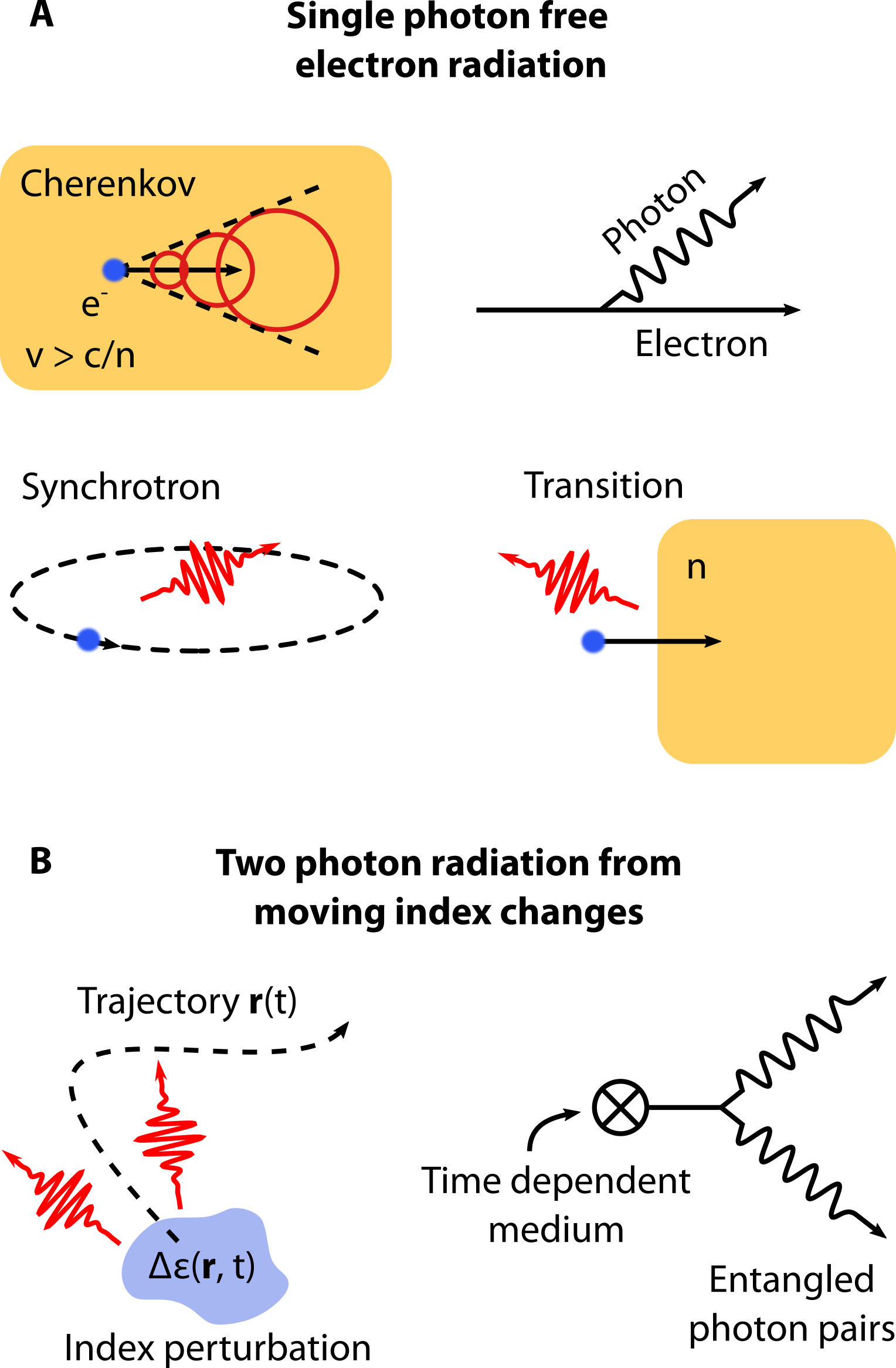}
    \caption{\textbf{Analogy between free electron radiation and quantum radiation from moving index perturbations.} (a) Free electrons can emit electromagnetic radiation through processes such as Cherenkov, synchrotron, and transition radiation. These processes are described to a high accuracy by classical electrodynamics, and in a quantum description are dominated by single photon emission processes, as depicted by the Feynman diagram on the top right. (b) In contrast, a moving index perturbation $\Delta\varepsilon$ on a trajectory $\rv(t)$ emits entangled photon pairs. The time dependence of the medium acts as a pump which creates nonclassical photon-pair states, as depicted in the Feynman diagram on the bottom right.}
    \label{fig:schematic}
\end{figure}

\section{Theoretical methods}

Our results are based on a Hamiltonian framework which describes two-photon emission in time-changing media. We consider a background structure (waveguide, photonic crystal, 2D material, etc.) which has a permittivity $\varepsilon_{\text{bg}}(\rv)$. A time-dependent perturbation is then applied to the material which causes the permittivity to undergo a change $\Delta\varepsilon(\rv, t)$, inducing a polarization $\Pv(\rv, t) = \varepsilon_0\Delta\varepsilon(\rv,t)\Ev(\rv, t)$. An interaction Hamiltonian $V(t) = -\varepsilon_0\int d^3r\, \Pv(\rv,t)\cdot\Ev(\rv,t)$ can then be written in terms of the material perturbation as
\begin{equation}
    V(t) = -\varepsilon_0 \int d^3r\,\Delta\varepsilon_{ij}(\rv, t)E_i(\rv,t) E_j(\rv,t),
    \label{eq:interaction_H}
\end{equation}
where we have used repeated index notation. Here, $\Ev(\rv, t) = \sum_n \sqrt{\frac{\hbar\omega}{2\varepsilon_0}}\left(\mathbf{F}_n(\rv)a_n e^{-i\omega_n t} + \mathbf{F}_n^*(\rv)a_n^\dagger e^{i\omega_n t}\right)$ is the interaction picture electric field operator, written in terms of creation and annihilation operators $a_n^{(\dagger)}$ for eigenmodes $\mathbf{F}_n(\rv)$ of the background structure. These eigenmodes satisfy the Maxwell equation $\nabla\times\nabla\times\mathbf{F}_n(\rv) =  \frac{\omega^2}{c^2}\varepsilon_{\text{bg}}(\rv)\mathbf{F}_n(\rv)$, and are normalized such that $\int d^3r\,\mathbf{F}_n^*(\rv)\cdot\varepsilon(\rv)\cdot \mathbf{F}_n(\rv) = 1$. Strictly speaking, this mode expansion assumes that the background structure has low dispersion and low loss. In the S.I., we outline a derivation in terms of macroscopic quantum electrodynamics (MQED) \cite{scheel2009macroscopic, rivera2020light}, which relaxes these assumptions. Additionally, we note that this framework assumes that the time-variation $\Delta\varepsilon(\rv, t)$ occurs over a region of the background structure which can be taken to be nondispersive. For details on what occurs when this assumption is relaxed, see \cite{sloan2020casimir}. 

The Hamiltonian (Eq. \ref{eq:interaction_H}) enables the photon field of the background structure $\varepsilon_{\text{bg}}$ to make transitions between initial and final states $\ket{i}$ and $\ket{f}$. To isolate two-photon emission processes, we take the initial state as vacuum ($\ket{i} = \ket{0}$), and the final state as a two photon state ($\ket{f} = \ket{m,n}$). The notation for the final state indicates that one photon is in mode $m$, and the other in mode $n$. In the absence of time dependence, such a process clearly does not conserve energy. However, the time dependence of the medium perturbation $\Delta\varepsilon$ acts as a classical pump field for entangled photon pairs so that energy is still conserved (similarly to the Hamiltonian theory of SPDC). The amplitude of emission is given by $S_{fi} = -\frac{i}{\hbar}\int dt\braket{f|V(t)|i}$. The total probability of this event is obtained by summing the amplitude from the S-matrix over all mode pairs $m$ and $n$ as $P = \frac{1}{2}\sum_{m,n}|S_{fi}|^2$, which gives the final result
\begin{equation}
    P = \frac{1}{8}\sum_{m,n}\omega_m\omega_n \left|\int d^3r\,\Delta\varepsilon_{ij}(\rv, \omega_n +\omega_m)F_{mi}^*(\rv)F_{nj}^*(\rv)\right|^2.
    \label{eq:master_probability_mode_expansion}
\end{equation}
Here, $\Delta\varepsilon(\rv, \omega) \equiv \int dt\,e^{i\omega t}\Delta\varepsilon(\rv, t)$ is the Fourier transform of the time varying index, and is \emph{not} related to dispersion. This probability can then be converted into transition rates, and provides the basis for our results. Additionally, the $S$-matrix encodes information about the quantum state of the emitted photon pairs, which can be used to analyze the quantum statistics of the emitted pairs.

\section{Two-photon Cherenkov radiation}
\label{sec:cherenkov}

\begin{figure*}[t]
    \centering
    \includegraphics[width=0.8\textwidth]{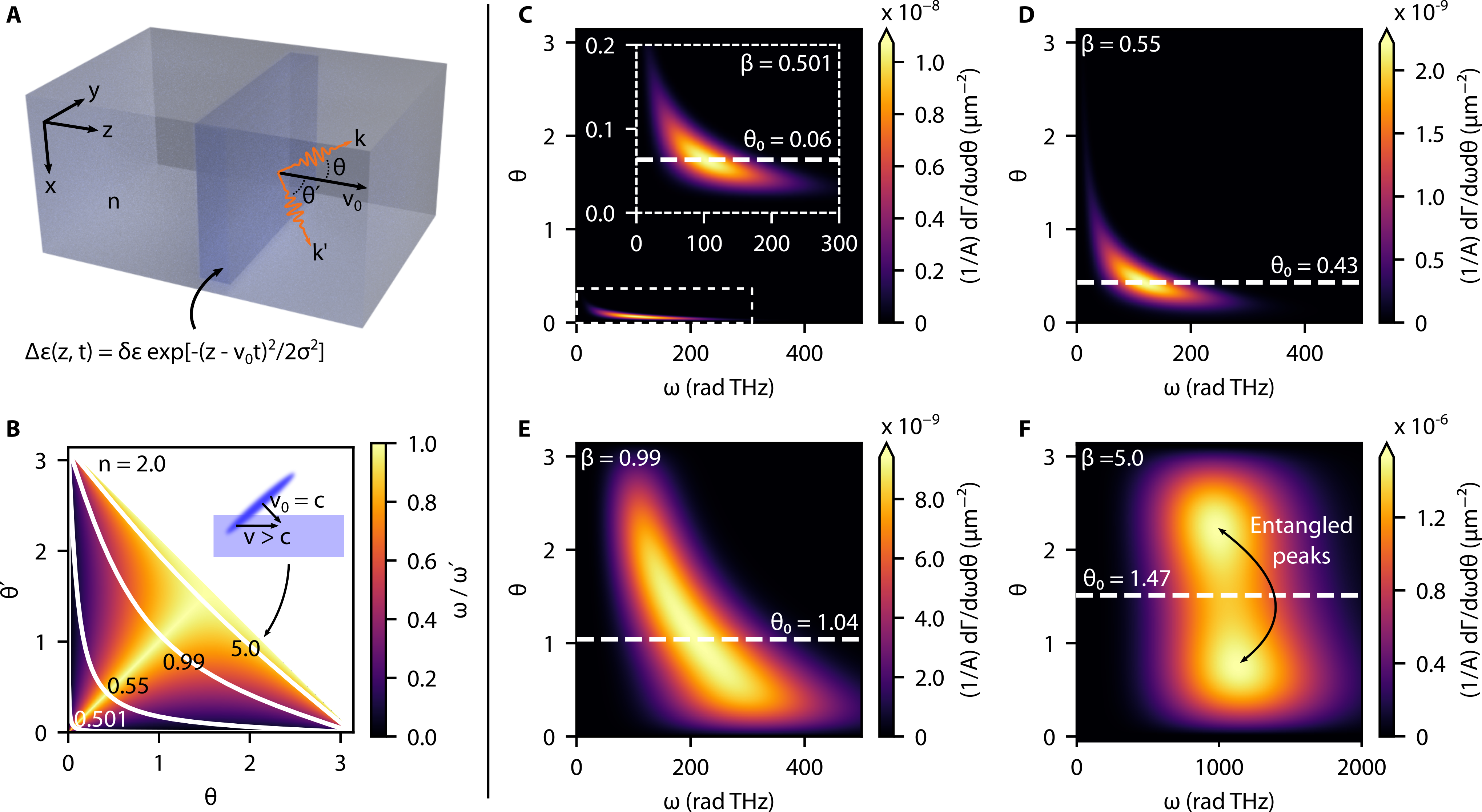}
    \caption{\textbf{Two-photon Cherenkov radiation by a superluminal index perturbation.} (a) Schematic of the geometry in which a wall with Gaussian profile (Eq. \ref{eq:superluminal_perturbation}) moves at velocity $v_0 = \beta c$, emitting entangled photon pairs with wavevectors $k$ and $k'$ in the background medium of index $n$. (b) Kinematics from Eq. \ref{eq:2ph_CR_kinematics}, which show the angular and spectral correlations of emitted photon pairs. Knowledge of one angle $\theta$ determines the angle of the other angle $\theta'$, shown by the contours for marked values of $\beta$. The inset shows how $\beta > 1$ can be achieved by a short light pulse incident on a nonlinear material at an angle, realizing truly superluminal behavior at the boundary. (c-f) Area-normalized angular frequency spectra $(1/A) d\Gamma/d\omega d\theta$ for index perturbations propagating at velocity $\beta c$ in a background index $n=2$, with $\sigma = 1\,\mu$m and $\delta\varepsilon/\varepsilon = 10^{-3}$. Velocities of $\beta=\{0.501, 0.55, 5.0\}$ are shown. Two-photon radiation onsets at the Cherenkov threshold $\beta > 1/n$, which is in this case at $\beta = 0.5$. As $\beta$ increases, the angular spread increases, deviating beyond the classical Cherenkov angle $\theta_0$ (marked with dashed lines). For a strongly superluminal pulse (f), higher frequencies are emitted, and at two angular peaks which depart from $\theta_0$. The angular correlations dictate that the photons at these two peaks are entangled with one another.}
    \label{fig:free_CR_differential_rate}
\end{figure*}

An important result of classical electrodynamics is the Cherenkov effect, which explains how a charge moving through a medium at constant speed emits light if it exceeds the phase velocity of light in the medium \cite{cherenkov1934visible, tamm1937coherent}. This phenomenon was first described in a homogeneous medium with constant refractive index, but applies far more generally \cite{kaminer2016efficient, luo2003cerenkov}. A nonlinear polarization in a medium can also behave effectively as a moving charge, emitting ``Cherenkov radiation'' when its velocity exceeds the phase velocity of a mode that it can couple to \cite{askaryan1962cerenkov, auston1984cherenkov, akhmediev1995cherenkov, vijayraghavan2013broadly}. Most of these processes have been realized in $\chi^{(2)}$ nonlinear materials, and are based on difference frequency generation (DFG) processes in which pump fields at two frequencies $\omega_1$ and $\omega_2$ combine to form a signal at $\omega_1 - \omega_2$ \cite{boyd2019nonlinear}. Consequently, these processes can only produce classical light (i.e. light in coherent states). 

Instead of an electron, our two-photon Cherenkov radiation relies on an index perturbation $\Delta\varepsilon$ which travels through a background medium at constant velocity. Photon pairs are emitted when this velocity is superluminal with respect to the phase velocity of light in the medium. We find that in a uniform dielectric, the two-photon nature of the process broadens the angular spectrum so that radiation can be produced at angles outside of the classical ``Cherenkov cone,'' even enabling backward two-photon Cherenkov radiation. In the Supplementary Information (S.I.) Section II, we provide details about how these concepts can also be applied to nanopolaritonic platforms for greater control and higher efficiency, using generation of graphene plasmon pairs as an example. We show that this plasmonic two-photon Cherenkov effect is more efficient than that in a uniform medium by 4 orders of magnitude.

To highlight the physical principles of the two-photon quantum Cherenkov phenomenon, we consider a homogeneous medium with index $n$. Then, we move a ``soft wall'' of index perturbation through the medium at velocity $v_0 = \beta c$ (Fig. \ref{fig:free_CR_differential_rate}a). The wall has an effective thickness $\sigma$ in the direction of propagation ($z$), and an area $A$ (which is large compared to the wavelength) in the transverse directions ($x, y$). We model this perturbation by setting
\begin{equation}
    \Delta\varepsilon(\rv, t) = \delta\varepsilon\,\exp\left[-\frac{(z - v_0 t)^2}{2\sigma^2}\right],
    \label{eq:superluminal_perturbation}
\end{equation}
which gives the boundary of the wall a Gaussian profile. When $\beta > 1/n$, pairs of photons can be emitted into the surrounding medium which has a dispersion relation $\omega_k = ck/n(\omega)$. In writing this, we assume that frequencies of interest are away from strong dispersion and absorption, so that $n(\omega)$ is slowly varying, and well described by $n = \sqrt{\varepsilon}$. The photons propagate with wavenumbers $k$ and $k'$, and with angles from the $z$-axis $\theta$ and $\theta'$ respectively. As a consequence of momentum conservation, the photons propagate in opposite directions in the $xy$ plane, with polar angles satisfying the constraint $\phi' = \phi + \pi$. We find a final differential decay rate per unit frequency and angle (normalized to the wall area $A$)
\begin{equation}
\begin{split}
    \frac{1}{A}\frac{d\Gamma}{d\omega d\theta} = \frac{v_0}{8\pi\omega_0^2}\left(\frac{\delta\varepsilon}{\varepsilon}\right)^2 &\left(\frac{n}{c}\right)^3 (\omega\omega')^2 e^{-2(\omega+\omega')^2/\omega_0^2} \\
    &\times\frac{\left(1 + (\cos\theta\cos\theta' - 1)^2\right)}{\left(\frac{\cot\theta}{n\beta} - \frac{1}{\sin\theta}\right)},
\end{split}
\label{eq:free_CR}
\end{equation}
where we have defined the frequency $\omega_0 = \sqrt{2}v_0/\sigma$, and by kinematic constraints, 
$\cos\theta' = (\omega/\omega' + 1)/n\beta - (\omega/\omega')\cos\theta$ and $\omega' = \omega(1 + n^2\beta^2 - 2n\beta\cos\theta)/(n^2\beta^2 - 1)$. For a relativistic free electron moving through a dielectric, Cherenkov radiation is a predominantly single-photon process. Consequently, classical electrodynamics describes experimental observations to a high accuracy. In contrast, the radiation described here is a two-photon process which is fundamentally quantum. As a consequence, the two emitted photons are entangled in frequency and momentum. Their angles of emission are subject to the constraint
\begin{equation}
    \sin\theta + \sin\theta' = n\beta \sin(\theta+\theta'),
    \label{eq:2ph_CR_kinematics}
\end{equation}
and their frequencies to the constraint $\omega\sin\theta = \omega'\sin\theta'$. The latter is a direct consequence of momentum conservation in the direction transverse to propagation, which arises from assuming that the area $A$ is large compared to the emitted wavelengths. In Fig. \ref{fig:free_CR_differential_rate}b, we see the contours defined by Eq. \ref{eq:2ph_CR_kinematics} for a background index $n=2$, and for several values of $\beta$. The colormap underneath the contour lines shows the ratio $\omega/\omega'$ between the two photon frequencies at each point $(\theta, \theta')$. Together, this information indicates the entanglement of the photon pair in both direction and frequency. 

Examining the angular and frequency distribution of two-photon radiation described by Eq. \ref{eq:free_CR} further reveals how this process departs from classical Cherenkov radiation. The frequency scale of the emitted radiation is set by both the velocity $v_0$ and width $\sigma$ of the perturbation. In particular, the frequency scale $\omega_0$ sets an exponential cutoff on the frequency sum $\omega+\omega'$ which can be emitted. Additionally, we see that higher frequencies are favored (far below the exponential cutoff), with $d\Gamma/d\omega d\theta \propto \omega^4$. Finally, we note that no radiation is produced unless the velocity exceeds the classical Cherenkov radiation condition $\beta > 1/n$. Figs. \ref{fig:free_CR_differential_rate}c-f shows the area-normalized spectrum $(1/A) d\Gamma/d\omega d\theta$ for a background index $n=2$, $\delta\varepsilon/\varepsilon = 10^{-3}$, and $\sigma = 1 \mu$m at different velocities $\beta$. Dashed white lines mark the classical Cherenkov angle $\theta_0 \equiv \cos^{-1}(1/\beta n)$. Since the background index is $n=2$, radiation onsets at $\beta = 0.5$. Just above threshold (Fig. \ref{fig:free_CR_differential_rate}c), the angular spectrum peaks sharply around the classical Cherenkov angle. As $\beta$ increases (Figs. \ref{fig:free_CR_differential_rate}d-f), so does the spread of the distribution in both angle and frequency. At $\beta = 0.99$, a substantial amount of the radiation goes backward ($\theta > \pi/2$). This feature is owed to the additional degree of freedom offered to the phase matching and momentum conservation conditions by the two-photon nature of this radiative process. If one photon is emitted at a higher angle and lower frequency, its partner photon can be emitted with higher frequency but lower angle so that the kinematic equations can still be satisfied. 

Although no physical object can exceed the speed of light, perturbations to the refractive index do not necessarily obey this constraint, leading to curious implications of our work. For example, consider a temporally short light pulse which is incident on a nonlinear interface at an angle. For angles sufficiently near normal incidence, the intersection of the pulse with the nonlinear material will create an index perturbation which propagates along the boundary with $v > c$ \cite{caloz2019spacetime}. This configuration is illustrated in the inset of Fig. \ref{fig:free_CR_differential_rate}b. We examine the consequences of this behavior by considering the geometry in Fig. \ref{fig:free_CR_differential_rate}a for $\beta = 5$ (Fig. \ref{fig:free_CR_differential_rate}f). The high velocity gives a cutoff frequency $\omega_0 \propto \beta$ which is several times higher than for $\beta > 1$. As a result, the emitted frequencies lie around 1 eV. Interestingly, this scenario gives rise to angular correlations which depart from behaviors seen for $\beta < 1$. The peak of the angular distribution no longer lies at the classical Cherenkov angle $\theta_0$, but rather occupies two peaks which are offset from $\theta_0$. There are, in some sense, two Cherenkov angles which mark the maxima of the distribution. Moreover, we see in the kinematic plot for this velocity (Fig. \ref{fig:free_CR_differential_rate}b) that as $\beta$ becomes large, the angular correlations approach the line $\theta + \theta' = \pi$; consequently, the ratio $\omega/\omega' \to 1$, indicating the photons are emitted near the same frequency. This means that a photon at one of the angular peaks is necessarily the entangled partner of that at the other peak. 

The ability to observe such a phenomenon hinges on the total rate of emission which can be detected. The total rate per area $\Gamma$ of the two-photon Cherenkov process is determined by integrating the angular spectral distributions $\Gamma = \int_0^{2\pi}d\theta \int_0^\infty d\omega\,(d\Gamma/d\omega d\theta)$. For $\beta = 0.55$, we have $\Gamma/A = 1.2 \times 10^{5}$ s$^{-1}\,\mu$m$^{-2}$. The very superluminal case ($\beta = 5$) exhibits a much stronger effect, with $\Gamma/A = 3.4 \times 10^{9}$ s$^{-1}\,\mu$m$^{-2}$. At these photon energies, this corresponds to an emitted power around 300 pW. We note that there is an inherent tradeoff in a system that creates superluminal perturbations, as an angle of incidence which gives a higher velocity also results in a larger effective value of $\sigma$. 

\section{Synchrotron radiation}

\begin{figure}[t]
    \centering
    \includegraphics[width=\linewidth]{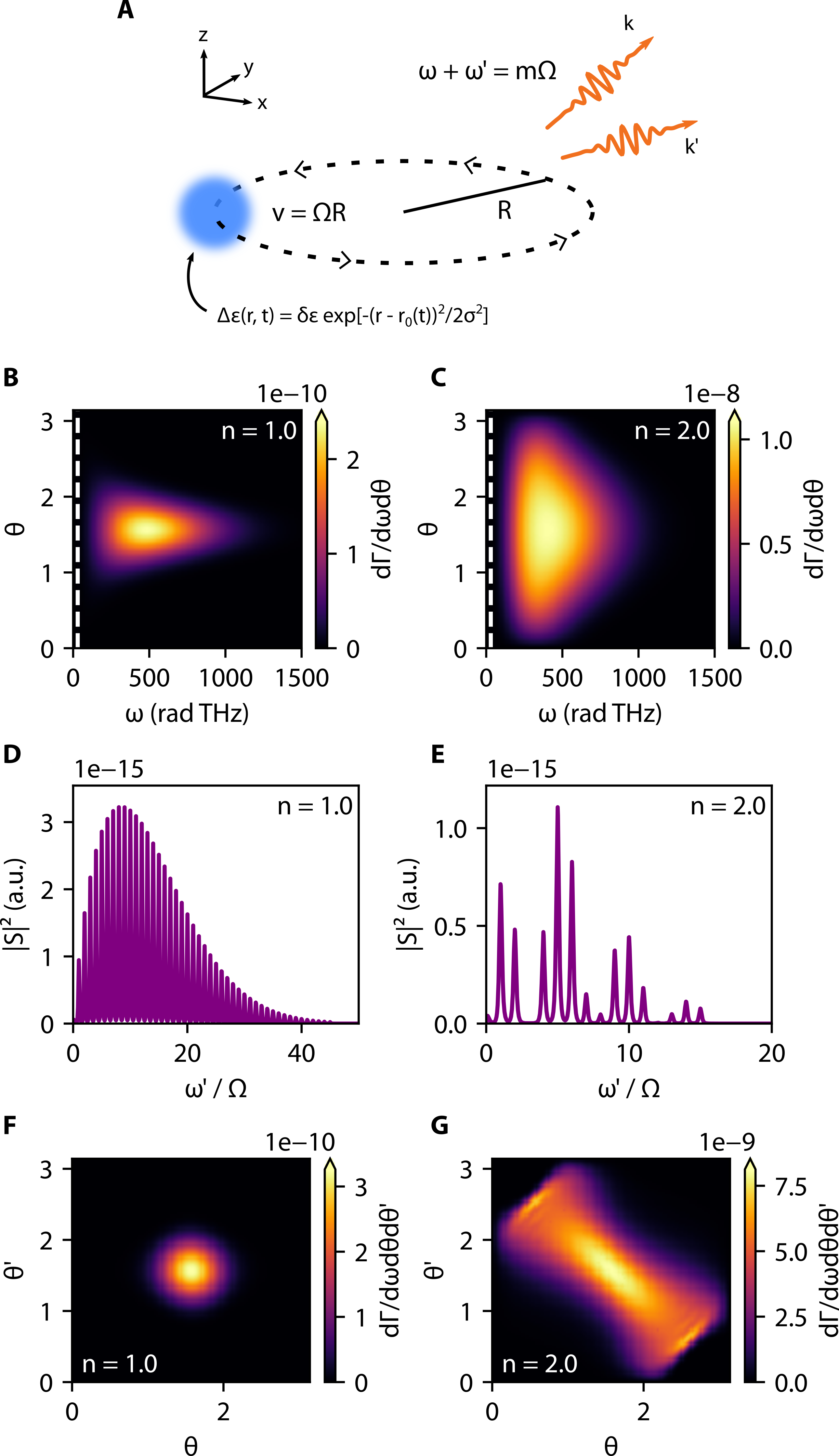}
    \caption{\textbf{Two-photon synchrotron radiation by an accelerating index perturbation.} (a) Schematic of index perturbation moving along a circular trajectory or radius $R$ with velocity $v = \Omega R$, where $\Omega$ is the angular rate of precession. Photon pairs are emitted with wavevectors $\kv$ and $\kv'$, and with frequencies which satisfy $\omega + \omega' = m\Omega$, where $m$ is an integer. (b, c) Angle and frequency spectrum of two-photon emission $d\Gamma/d\omega d\theta$ for $\sigma = 1\,\mu$m and $v = c$, $R$ = 10 $\mu$m, $\delta\varepsilon/\varepsilon = 10^{-3}$ for $n = \{1, 2\}$. (d, e) Frequency correlations at $\theta = \theta' = \pi/2$ and $\phi = \phi'$ when one photon $\omega = 500$ rad-THz is measured; the other frequency is forced to lie on a comb $\omega' = m\Omega - \omega$. (f, g) Angular correlations shown through $d\Gamma/d\omega d\theta d\theta'$ as a function of $\theta, \theta'$ for $\omega = 500$ rad-THz.}
    \label{fig:synch}
\end{figure}

The two-photon Cherenkov radiation described above originates from superluminal phase matching. We now show that an accelerating index perturbation also radiates. As an example, we show that when an index perturbation traverses a circular trajectory (Fig. 4a), photon pairs are emitted in a manner which bears some similarity to synchrotron radiation \cite{sokolov1966synchrotron, wiedemann2003synchrotron, kunz1974synchrotron, kim1989characteristics, ginzburg1965cosmic, schwinger1998classical}. In a synchrotron, an electron is accelerated to near-light speeds along a kilometer-scale circular trajectory. As a result, the accelerating particle can emit very high harmonics of its angular frequency $\Omega$, making synchrotrons important sources of X-ray photons. In the two-photon optical ``synchrotron'' we propose here, the radiation wavelength is set primarily by the size of the perturbation. However, the harmonic nature of the radiation persists in the frequency-frequency correlations of the photons pairs.

To model the two-photon synchrotron effect, we consider an index perturbation (Fig. \ref{fig:synch}a) $\Delta\varepsilon(\rv, t) = \delta\varepsilon \,e^{-\frac{1}{2\sigma^2}(\rho^2 + z^2 + R^2)}e^{\frac{R\rho}{\sigma^2}\cos(\theta - \Omega t)}$, where $(\rho, \theta, z)$ are cylindrical coordinates, $\Omega$ is the angular frequency of precession, $\sigma$ is the Gaussian width, and $R$ is the radius of the circle. As with the Cherenkov process, we assume that the perturbation is applied to a background permittivity $\varepsilon$. The emitted photon frequencies must sum to an integer multiple $m$ of the precession frequency, as $\omega + \omega' = m\Omega$. The rate of pair generation per unit angle $\theta$ and frequency $\omega$ into each harmonic $m$ is given as:
\begin{equation}
\begin{split}
       \frac{d\Gamma_m}{d\omega d\theta} = &\left(\frac{\delta\varepsilon}{\varepsilon}\right)^2 \frac{\sigma^6 n^6 \omega^3 (m\Omega - \omega)^3\sin\theta}{16\pi c^6} \\
       &\times\int d\Omega' \, \left(1 + (\hat{\kv}\cdot\hat{\kv}')^2\right)e^{-\sigma^2\left(\kv+\kv'\right)^2} J_m^2(KR)
\end{split}
\label{eq:synch}
\end{equation}
Here,  $K = \sqrt{(k_x + k_x')^2 + (k_y + k_y')^2}$ is the net in-plane wavevector of emission, and $J_m$ is the $m$-th Bessel function. 

Similarly to the quantum Cherenkov radiation, we can understand the properties of this nanophotonic synchrotron radiation in terms of its spectrum, as well as its angular and spectral correlations. In two different background indices ($n=1, 2$), the total spectrum $d\Gamma/d\omega d\theta$ summed over all harmonics $m$ is smooth, and centered about $\theta=\pi/2$ (Figs. \ref{fig:synch}b,c). A higher background index results in a larger angular spread, similar to classical Cherenkov radiation where a higher index causes phase-matching to occur at larger angles. Increasing the background index also results in a substantially higher output power as can be seen in the magnitude of overall photon rate. From Eq. \ref{eq:synch}, we see that the emitted photon rate at a given frequency $\omega$ depends directly on how the emitted wavenumber $n\omega/c$ compares to the length scale set by $\sigma$ (and likewise for the other photon at $n(m\Omega - \omega)/c$). In this example, $\sigma = 1\, \mu$m sets the frequency scale of emitted radiation around $500$ rad-THz. 

We also highlight the frequency correlations (Figs. \ref{fig:synch}d,e) which emerge by fixing the frequency of $\omega'$ and looking at the amplitude of the $S$ matrix elements as $\omega$ is allowed to vary. Due to the constraint between frequencies, $\omega$ must lie in a comb spaced by $\Omega$. If one photon $\omega$ is measured, then the other remains in a quantum state of the form $\ket{\psi} = \sum_m c_m \ket{\omega - m\Omega}$, with coefficients $c_m$ determined through the $S$-matrix. When the index is raised, the perturbation becomes superluminal in addition to its acceleration, and interference fringes emerge in the envelope of angular correlations (Fig. \ref{fig:synch}e). Thus, we see that the quantum state created by heralding can be controlled through the parameters of the nanophotonic synchrotron. Such a source could present new opportunities for quantum information, or as a quantum probe of atomic, molecular, or solid-state systems. Varying the background index also influences the angular correlations of photon pairs (Figs. \ref{fig:synch}f,g). For $n=1$, emission at $\theta = \theta' = \pi/2$ is strongly preferred. As $n$ increases, the angular distributions change substantially. We thus see that by changing basic parameters of the system, or the observation, one can create a rich variety of correlations in angle and frequency. 

\section{Experimental outlook}

We now provide more detailed information about what is needed in order to experimentally realize the two-photon mechanisms described above. To realize the nanophotonic ``synchrotron,'' light in a Kerr medium could be guided along a curved trajectory. One particularly interesting system for this could be temporal solitons (with FWHM as low as 30 fs) in ring resonators (with radius as low as $R = 10\,\mu$m) \cite{kippenberg2018dissipative, brasch2016photonic, herr2014temporal, herr2014mode, pfeiffer2017octave}, which have been used for on-chip frequency comb generation. In these systems, nonlinearity and dispersion balance each other out, enabling a micron-scale pulse of light to preserve its shape while propagating around a ring resonator. Due to the Kerr nonlinearity, this pulse of light is accompanied by an index change similar to that depicted in Fig. \ref{fig:synch}a. The emitted frequencies will depend on the size of this index perturbation in its propagation and transverse directions. Depending on the parameters, photon pairs could be emitted directly into the waveguide, into the far-field, or both. Using the parameters from Fig. \ref{fig:synch} in a background index $n=2$, we estimate that photon pairs are produced at a rate of $\Gamma \approx 10^{4}$ s$^{-1}$. One particularly attractive aspect of this platform is that it could eventually lead to the generation of entangled frequency combs on-chip. Other ways to achieve the acceleration of light pulses include metasurfaces \cite{henstridge2018synchrotron}, self-accelerating beams \cite{kaminer2011self, schley2014loss, dolev2012experimental}, topological edge-state solitons \cite{leykam2016edge}, and soliton pairs which spiral around each other along a helical trajectory \cite{shih1997three, buryak1999induced}.

Superluminal and accelerating index perturbations could also be created with short pulses of light colliding with Kerr nonlinear structures. A concept for realizing superluminal perturbations was discussed in the context of homogeneous medium Chereknov radiation (Figs. \ref{fig:free_CR_differential_rate}b,f). This configuration could be realized with an intense femtosecond \cite{akhmanov1988optics, rulliere2005femtosecond} or attosecond \cite{paul2001observation, sansone2006isolated} pulse which impinges on a nonlinear interface. For the parameters used in Fig. \ref{fig:free_CR_differential_rate}f, photon pairs are produced at a rate of $\Gamma/A = 3.4 \times 10^{9}$ s$^{-1}\,\mu$m$^{-2}$. An imaging system could then be used to collect the photons over some frequency range to measure their angles of emission. By varying the angle of incidence of the pulse, one controls the velocity of the perturbation at the interface, allowing access to different regimes of angular emission. We note that for a pulse of fixed width in its propagation direction, an angle which yields a higher velocity also yields an effective width which is smaller by the same proportion, leaving $\omega_0 \propto \beta/\sigma$ fixed. This system thus presents intriguing opportunities to study the physics of effective ``tachyons,'' hypothetical particles that travel faster than light \cite{feinberg1970particles}.

As alluded to previously, this configuration could also be suitable for creating the type of index perturbation for generating graphene plasmons, or other surface excitations (discussed in detail in the S.I.). One could imagine depositing graphene over a nonlinear waveguide structure, through which a soliton pulse is propagated, generating plasmon pairs on the graphene. The ability to observe entangled plasmon production from such a system is determined by the strength of the emission, as well as the capability of detection, on a realizable platform. For example, if the index perturbation comes from a soliton ($W = 10 \mu$m) propagating through a Kerr nonlinear substrate near graphene with parameters as described in supplementary Fig. S1, graphene plasmon pairs are produced at a rate exceeding $10^9$/s. This corresponds to a power of around 100 pW, which at these near IR frequencies, should enable correlation measurements at the single-plasmon level.

In addition to creating linear perturbations, this concept can be extended to different geometries. For example, colliding an ultrashort pulse with a nonlinear structure shaped like a spring would create an index perturbation which traverses a helical trajectory, thus realizing acceleration with a periodicity as described in the ``synchrotron'' geometry (see Supplemental Fig. S2). Other related systems which may provide opportunities for inducing controlled index perturbations of optical size include subdiffractional plasmon-solitons \cite{feigenbaum2007plasmon, davoyan2009self, kuriakose2020nonlinear, nesterov2013graphene}, spatiotemporal solitons \cite{liu1999generation, malomed2005spatiotemporal}, and so-called ``light-bullets'' \cite{panagiotopoulos2015super, abdollahpour2010spatiotemporal}.


\section{Conclusion}

We have presented a new concept for two-photon emission based on moving index perturbations along a controlled trajectory. Our work points toward a paradigm where the spatial trajectory of such pulses can influence the spectrum, direction, and entanglement of emitted radiation. The nanophotonic ``synchrotron'' could enable the generation of heralded single-photon frequency combs, which could pave the way toward developments in quantum optics, metrology, and information processing. Future work on this topic could explore the pair emission created by index perturbations moving in a trajectory which accelerates linearly, or impinges on an interface between two materials, leading to respective analogs of two-photon analogs of Unruh and transition radiation. Different photonic structures may also suggest new methods for shaping the emitted radiation. For example, time-modulated photonic crystal structures \cite{skorobogatiy2000rigid} may present particularly interesting opportunities for controlling these time-dependent quantum effects, due to their controllable density of states, along with constantly improving nanofabrication techniques. It would also be potentially interesting to develop a detailed account of the quantum statistics, squeezing, and entanglement of light which are generated through these systems. Broadly, we anticipate our work should serve as a starting point for using spatiotemporal control over light pulses to create quantum states of light.


\begin{acknowledgements}
    The authors thank Dr. Yannick Salamin and Prof. Ido Kaminer for helpful discussions. This material is based upon work supported in part by the Defense Advanced Research Projects Agency (DARPA) under Agreement No. HR00112090081. This work is supported in part by the U.S. Army Research Office through the Institute for Soldier Nanotechnologies under award number W911NF-18-2-0048. This material is also based upon work supported by the Air Force Office of Scientific Research under the award number FA9550-20-1-0115. J.S. was supported in part by Department of Defense NDSEG fellowship No. F-1730184536. N.R. was supported by Department of Energy Fellowship DE-FG02-97ER25308. 
\end{acknowledgements}

\bibliography{vacuum}
\newpage
\onecolumngrid
\appendix 

\newpage
\end{document}


\rmfamily

\title{Supplementary Information: \\
Two-photon emission from superluminal and accelerating index perturbations}
\author{Jamison Sloan$^{1\dagger}$, Nicholas Rivera$^{2}$, John D. Joannopoulos$^{2}$, and Marin Solja\v{c}i\'{c}$^{2}$}

\affiliation{$^{1}$ Department of Electrical Engineering and Computer Science, Massachusetts Institute of Technology, Cambridge, MA 02139, USA \\
$^{2}$ Department of Physics, Massachusetts Institute of Technology, Cambridge, MA 02139, USA
$\dagger$ Corresponding author e-mail: jamison@mit.edu}

\noindent

\maketitle
\tableofcontents
\newpage 

In this supplement, we provide additional details in support of the main text of this work. In Section \ref{sec:general_theory}, we outline the theoretical framework used to obtain our results. In Section \ref{sec:plasmon_cherenkov}, we provide additional results about the production of graphene plasmon pairs from a fast index perturbation moving above the surface. In Section \ref{sec:derivations}, we outline the derivations used to obtain results for specific example systems discussed in this work. Section \ref{sec:green_function} provides a summary of properties of Maxwell Green's functions in layered media, which are used in the derivations for the two-plasmon Cherenkov effect in graphene. 

\section{General theoretical formalism}
\label{sec:general_theory}

Here, we elaborate on the theoretical foundations of two-photon emission in time-dependent media which are summarized in the Methods section of the main text. More details about the theoretical foundations of two-photon emission in general optical media can be found in \cite{sloan2020casimir}.

\subsection{Mode expansion}

We begin by deriving an expression for the rate of two-photon emission using an eigenmode expansion for the quantized electric field operator. We apply perturbation theory to calculate elements of the $S$-matrix, which yields emission probabilities, as well as the quantum state of the system after emission. The Hamiltonian of the bare field is given by
\begin{equation}
    H = \sum_n \hbar\omega_n\left(a_n^\dagger a_n + \frac{1}{2}\right)
\end{equation}
where $a_n$ is the annihilation operator for a mode with label $n$. In space, the modes are assumed to be orthogonalized under the inner product $\int d^3r\,\mathbf{F}_m^*(\rv)\cdot\varepsilon_{\text{bg}}(\rv)\cdot\mathbf{F}_n(\rv) = \delta_{mn}$. Here, $\varepsilon_{\text{bg}}(\rv)$ is the permittivity of the background structure, which is assumed here to be lossless and nondispersive. Then, the interaction picture  electric field operator is given via a sum over eigenmodes $\mathbf{F}_n(\rv)$ 
\begin{equation}
    \Ev(\rv, t) = \sum_n \sqrt{\frac{\hbar\omega}{2\epsilon_0}}\left(\mathbf{F}_n(\rv)a_n e^{-i\omega_n t} + \mathbf{F}_n^*(\rv)a_n^\dagger e^{i\omega_n t}\right).
\end{equation}
The interaction Hamiltonian, written in the interaction picture, is given by 
\begin{equation}
    V(t) = -\varepsilon_0 \int d^3r\,\Delta\varepsilon_{ij}(\rv, t)E_i(\rv, t) E_j(\rv,t),
    \label{eq:interaction_H}
\end{equation}
where we have used index notation. We consider initial states which are the vacuum, and final states which have two photons. Thus we write $\ket{i} = \ket{0}$, and $\ket{f} = \ket{m,n}$, where $m, n$ are the mode labels. If the interaction Hamiltonian were time-independent, such a process would not be possible, as the initial and final states clearly do not conserve energy. However, the time dependence of the medium lifts this constraint, and the medium change $\Delta\varepsilon$ can act as a classical pump field. The transition amplitude at leading order in perturbation theory is given by
\begin{equation}
    S_{fi}(t) = -\frac{i}{\hbar}\int_{-\infty}^t dt'\braket{f|V(t')|i}.
\end{equation}
Plugging in the initial and final states defined above, we find that
\begin{equation}
    S_{fi}(t) = \frac{i}{2}\sqrt{\omega_n\omega_m} \int d^3r \int_{-\infty}^t dt'\, \Delta\varepsilon_{ij}(\rv,t')e^{i(\omega_m  +\omega_n)t'}F_{mj}^*(\rv)F_{ni}^*(\rv)
\end{equation}
In order to study the emission of asymptotic states which would be detected away from the source, we can take $t\to\infty$. This allows us to group the time integral, the exponential, and $\Delta\varepsilon$ as the Fourier transform of the change in susceptibility:
\begin{equation}
    \Delta\varepsilon(\rv, \omega) \equiv \int_{-\infty}^\infty dt\, e^{i\omega t}\Delta\varepsilon(\rv, t).
    \label{eq:permittivity_ft}
\end{equation}
For clarity, we note that this quantity is \emph{not} related to the dispersion of the modulated material, which is assumed here not to be significant. With Eq. \ref{eq:permittivity_ft}, the matrix element as $t \to \infty$ simplifies to
\begin{equation}
    S_{fi}^{t\to\infty} = \frac{i}{2}\sqrt{\omega_n \omega_m}\int d^3r\,\Delta\varepsilon_{ij}(\rv, \omega_n + \omega_m)F_{mj}^*(\rv)F_{ni}^*(\rv).
    \label{eq:s_matrix}
\end{equation}
To find the total probability $P$ from this matrix element, we take the modulus squared and sum over final photon states $m$ and $n$:
\begin{equation}
    P = \frac{1}{2}\sum_{m,n}|S_{fi}|^2 = \frac{1}{8}\sum_{m,n}\omega_m\omega_n \left|\int d^3r\,\Delta\varepsilon_{ij}(\rv, \omega_n + \omega_m)F_{mj}^*(\rv)F_{ni}^*(\rv)\right|^2,
    \label{eq:master_probability_mode_expansion}
\end{equation}
where the factor of two has come from the indistinguishability of photons labeled with $m$ and $n$. This is the result shown in Eq. 2 of the main text.

Critically, we see that the strength of the matrix element for two particular modes $m$ and $n$ depends on $\Delta\varepsilon$ evaluated at the sum of frequencies $\omega_m + \omega_n$. In this sense, $\Delta\varepsilon(\rv, \omega)$ directly controls the emitted spectrum. In our examples discussed in the main text, we see clearly how the spatial profile of $\Delta\varepsilon(\rv, \omega)$ leads directly to frequency and momentum phase-matching rules for the emitted photon pairs. 

\subsection{Macroscopic QED}

While the mode expansion version detailed above works well in systems which can be idealized as lossless and dispersionless, treating problems of emission in vacuum in arbitrary environments requires the formalism of macroscopic quantum electrodynamics (MQED) \cite{scheel2009macroscopic, rivera2020light}. In this framework, the Hamiltonian of the bare electromagnetic field is 
\begin{equation}
    H_{\text{EM}} = \int_0^\infty d\omega \int d^3r\, \hbar\omega\, \mathbf{f}^\dagger(\rv,\omega)\cdot\mathbf{f}(\rv,\omega),
\end{equation}
where $\mathbf{f}^{(\dagger)}(\rv,\omega)$ is the annihilation (creation) operator for a quantum harmonic oscillator at position $\rv$ and frequency $\omega$. In such a medium, the electric field operator in the interaction picture is given as 
\begin{equation}
    \Ev(\rv, t) = i \sqrt{\frac{\hbar}{\pi\varepsilon_0}} \int_0^\infty d\omega \frac{\omega^2}{c^2} \int d^3r' \sqrt{\Im \varepsilon_{\text{bg}}(\rv', \omega)} \left(G(\rv,\rv',\omega)\mathbf{f}(\rv',\omega)e^{-i\omega t} - \text{h.c.}\right).
\end{equation}
Here, $G(\rv,\rv',\omega)$ is the electromagnetic Green's function of the background which satisfies $\left(\curl\curl - \varepsilon_{\text{bg}}(\rv,\omega)\frac{\omega^2}{c^2}\right)G(\rv,\rv',\omega) = \delta(\rv-\rv')I$, where $I$ is the $3\times 3$ identity matrix. Additionally, we assume for generality that the modulated material can also be dispersive, and is described through a generalized linear response function $\Delta\varepsilon(\rv, t, t')$ (see \cite{sloan2020casimir} for details), so that the interaction Hamiltonian becomes
\begin{equation}
    V(t) = -\varepsilon_0 \int d^3r\,dt'\,\Delta\varepsilon_{ij}(\rv, t, t')E_j(\rv,t') E_i(\rv,t).
\end{equation}
We now take the initial and final states of the scattering problem to to be $\ket{i} = \ket{0}$ and $\ket{f} = \ket{m\rv_1\omega_1,n\rv_2\omega_2}$. Then the scattering matrix element is given as
\begin{align}
    S_{fi} &= -\frac{i}{\hbar}\int_{-\infty}^\infty dt \braket{f|V(t)|i} \\
    &= -\frac{i(\omega_1\omega_2)^2}{\pi c^4}\int d^3x \int_{-\infty}^\infty dt\,dt'\,e^{i\omega_1 t'}e^{i\omega_2 t} \sqrt{\Im \varepsilon_{\text{bg}}(\rv_1,\omega_1) \Im \varepsilon_{\text{bg}}(\rv_2,\omega_1)} \\
    &\hspace{2cm}\times\Delta\varepsilon_{ij}(\xv, t, t') G_{jm}^*(\xv, \rv_1, \omega_1)G_{in}^*(\xv, \rv_2, \omega_2).
\end{align}
From the S-matrix, we can sum over final states to obtain the probability of two-photon emission:
\begin{align}
    P &= \frac{1}{2}\sum_{m,n} \int d^3r_1\,d^3r_2\int_0^\infty d\omega_1\,d\omega_2\,|S_{fi}|^2 \\
    &= \frac{1}{2}\sum_{m,n} \int d^3r_1\,d^3r_2\,\int_0^\infty d\omega_1\,d\omega_2 \Bigg|\frac{(\omega_1\omega_2)^2}{\pi c^4}\int d^3x\, \sqrt{\Im \varepsilon_{\text{bg}}(\rv_1,\omega_1) \Im \varepsilon_{\text{bg}}(\rv_2,\omega_1)} \\
    &\hspace{3cm}\Delta\varepsilon_{ij}(\xv, \omega_2, -\omega_1) G_{jm}^*(\xv, \rv_1, \omega_1)G_{in}^*(\xv, \rv_2, \omega_2)\Bigg|^2.
\end{align}
Here, we have expressed $\Delta\varepsilon(\rv,t,t')$ in terms of its two-time Fourier transform, which we define as 
\begin{equation}
    \Delta\varepsilon(\rv, \omega, \omega') \equiv \int_{-\infty}^\infty dt\,dt'\, e^{i\omega t}\Delta\varepsilon(\rv, t, t')e^{-i\omega't'}.
\end{equation}
Finally we make use of the following identity \cite{scheel2009macroscopic} to simplify the result:
\begin{equation}
    \frac{\omega^2}{c^2}\int d^3x \Im \varepsilon_{\text{bg}}(\xv, \omega)G(\rv,\xv,\omega)G^\dagger(\rv',\xv, \omega) = \Im G(\rv, \rv', \omega)
\end{equation}
R elabeling variables, we can write a final expression for the probability as
\begin{equation}
\begin{split}
    P = \frac{1}{2\pi^2 c^4}\int_0^\infty &d\omega\,d\omega' \int d^3r\,d^3r'\,(\omega\omega')^2 \\
    &\times\tr\left[\Delta\varepsilon(\rv, \omega,-\omega')\Im G(\rv,\rv',\omega')\Delta\varepsilon^\dagger(\rv', \omega,-\omega')\Im G(\rv',\rv,\omega)\right].
\end{split}
\label{eq:master_prob}
\end{equation}
We note that in the limit of low loss in the Maxwell equations, the Imaginary part of the Green's function can be written in terms of the eigenmodes as 
\begin{equation}
    \Im G(\rv, \rv', \omega) = \pi c^2 \sum_n \frac{1}{2\omega_n}\mathbf{F}_n^*(\rv) \mathbf{F}_n(\rv')\delta(\omega - \omega_n).
\end{equation}
Using this correpsondence, Eq. \ref{eq:master_probability_mode_expansion} can be converted into Eq. \ref{eq:master_prob}, establishing a firm connection between the eigenmode expansion and the MQED approach. We briefly mention two important limits of Eq. \ref{eq:master_prob}. If the modulation to the material takes place in a region which can be regarded as nondispersive, then $\Delta\varepsilon(\rv, \omega, -\omega') = \Delta\varepsilon(\rv, \omega + \omega')$, where the latter form is that described by Eq. \ref{eq:permittivity_ft}. In this case, we have
\begin{equation}
\begin{split}
        P = \frac{1}{2\pi^2 c^4}\int_0^\infty &d\omega\,d\omega' \int d^3r\,d^3r'\,(\omega\omega')^2 \\
    &\times\tr\left[\Delta\varepsilon(\rv, \omega+\omega')\Im G(\rv,\rv',\omega')\Delta\varepsilon^\dagger(\rv', \omega+\omega')\Im G(\rv',\rv,\omega)\right].
\end{split}
\end{equation}
This enables one to describe a situation where a nondispersive modulation emits into a dispersive and lossy region nearby, which is the case for the graphene plasmon system discussed in Section \ref{sec:plasmon_cherenkov}. The other important limit is when the modulation is assumed to be isotropic, in which case $\Delta\varepsilon_{ij}(\rv, \omega, -\omega') = \delta_{ij}\Delta\varepsilon(\rv,\omega, -\omega')$. In this case, we have
\begin{equation}
\begin{split}
    P = \frac{1}{2\pi^2 c^4}\int_0^\infty &d\omega\,d\omega' \int d^3r\,d^3r'\,(\omega\omega')^2\Delta\varepsilon(\rv, \omega,-\omega')\Delta\varepsilon^\dagger(\rv', \omega,-\omega') \\
    &\times\tr\left[\Im G(\rv,\rv',\omega')\Im G(\rv',\rv,\omega)\right].
\end{split}
\end{equation}

\section{Cherenkov emission into graphene plasmons}
\label{sec:plasmon_cherenkov}

In a homogeneous medium of constant index $n$, interactions between the time-variation $\Delta\varepsilon$ and the electric field $E$ are quite weak. The coupling strength in the interaction Hamiltonian (Eq. \ref{eq:interaction_H}) is, roughly speaking, proportional to the magnitude of $\Delta\varepsilon$ as well as the fluctuations of the vacuum electromagnetic field $\Braket{E^2}$ over the region of index perturbation. A key paradigm in nanophotonics is the ability to engineer electromagnetic excitations which have much higher field strengths at the single photon level. These techniques have been applied to enhance spontaneous emission \cite{purcell1946purcell}, Casimir forces \cite{rodriguez2011casimir}, near-field radiative heat transfer \cite{volokitin2007near}, and more. One particularly interesting system for realizing these effects are thin materials which support highly confined polaritons \cite{atwater2007promise,basov2016polaritons,basov2017towards,low2017polaritons,iranzo2018probing,ni2018fundamental, caldwell2013low, caldwell2014sub, caldwell2015low}, thus enabling strong light-matter interactions.

\begin{figure*}[t]
    \centering
    \includegraphics[scale=0.9]{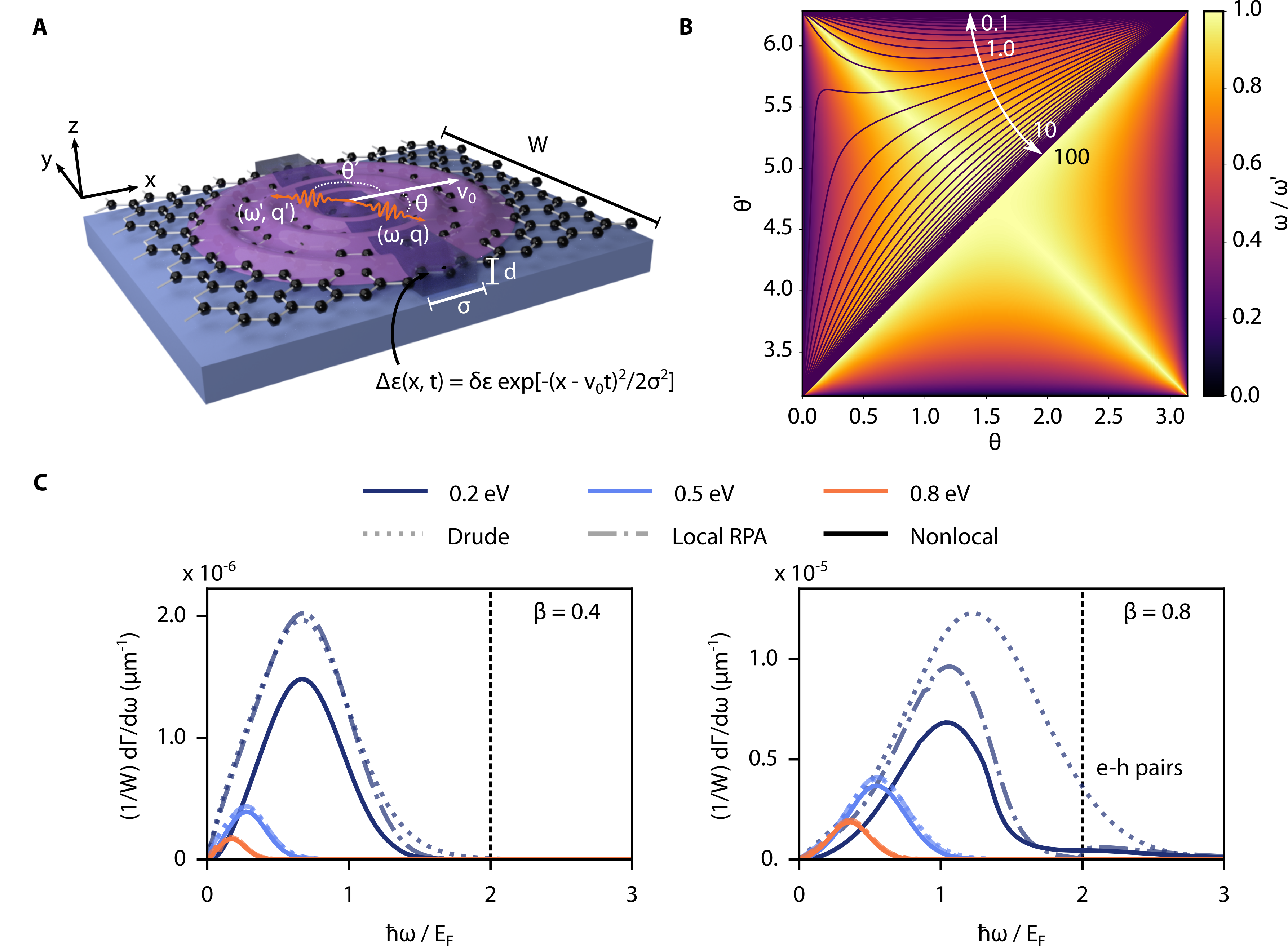}
    \caption{\textbf{Emission of entangled plasmon pairs by moving index perturbations near graphene.} (a) Schematic of an index perturbation $\Delta\varepsilon$ moving at constant velocity $v_0$ near a sheet of graphene. (b) Angular correlation of photon pairs emitted in lossless Drude model. Contours show values of constant $\hbar\omega\beta/(4\alpha E_{\text{F}})$, where $\alpha$ is the fine structure constant. The ratio between frequencies $\omega/\omega'$ is shown on the colorbar below, indicating at any point on a contour how one frequency relates to the other. (c) Frequency distribution $(1/W) d\Gamma/d\omega$ of emitted plasmon pairs for an index perturbation $\delta\varepsilon/\varepsilon = 10^{-3}$ of velocity $\beta c$ which has dimensions $\sigma = 400$ nm, $d = 1\,\mu$m. Graphene with three different doping levels $E_\text{F} = \{0.2, 0.5, 0.8\}$ eV are shown, and each with three different conductivity models. A relaxation constant $\tau = 10^{-14}$ s is used, and dispersion of $\tau$ may be taken into account using the methods provided in \cite{jablan2009plasmonics}. The electron-hole pair excitation region at $\hbar\omega = 2E_{\text{F}}$ is marked with a dotted line. Nonlocal corrections are small when the frequency spectrum does not substantially enter this region. At high velocity and low doping levels (right panel for 0.2 eV), the nonlocal corrections are substantial, and account for the production of electron-hole pairs.}
    \label{fig:plasmon_CR}
\end{figure*}

To demonstrate that this concept generalizes to our two-photon generation mechanism, we show how a fast index perturbation near a sheet of graphene \cite{jablan2009plasmonics, koppens2011graphene} produces entangled pairs of graphene plasmons with much higher efficiencies than in a homogeneous medium. While we focus on the graphene example, this concept readily generalizes to other planar systems (e.g. phonon-polaritons, exciton-polaritons, magnon-polaritons). We assume that a substrate experiences a gaussian perturbation, with a direction of propagation $x$ (Fig. \ref{fig:plasmon_CR}a). Additionally, the soft wall is given a width $W$ in the $y$ direction, which is assumed to be large compared to the relevant wavelengths, and assumed to extend uniformly over $0 < z < d$. Mathematically, this is defined as 
\begin{equation}
    \Delta\varepsilon(x, t) = \delta\varepsilon\,\exp\left[-\frac{(x - v_0 t)^2}{2\sigma^2}\right],
\end{equation}
for $0 < z < d$, and zero otherwise. The width of the wall in the propagation direction is denoted by $\sigma$, similarly to the free-space Cherenkov emission discussed in the main text. While we depict the perturbation to be free standing, this would in practice occur in a nonlinear substrate or superstrate, leading to effects similar in magnitude. This index perturbation then travels parallel to the graphene surface at some fraction the speed of light $\beta$. In the case of a uniform medium discussed in the main text, the ``moving wall'' of permittivity variation emits photon pairs into the surrounding space. Here, the relevant electromagnetic excitations are surface plasmon polaritons. Consequently, an index perturbation with sufficiently high velocity will emit pairs of entangled plasmons into the graphene sheet. Since the velocity of graphene plasmons can easily go below $c/100$, the superluminal condition is easier to reach than in a homogeneous medium. Using our formalism, we obtain a general expression (see Supplementary Section \ref{subsec:plasmon_derivation} for details) for the two-plasmon emission rate per unit frequency $\omega$ and in-plane angle $\theta$:
\begin{equation}
\begin{split}
    \frac{1}{W}\frac{d\Gamma}{d\omega d\theta} &= \frac{\beta c}{(2\pi)^3} \frac{\delta\varepsilon^2}{\omega_0^2}\int d\omega' \int dq\,q^2 q'\, e^{-2(\omega+\omega')^2/\omega_0^2} \\
    &\times [g(q+q')]^2 (\hat{q}\cdot\hat{q}' - 1)^2 \Im r_p(\omega, q) \Im r_p(\omega', q').
\end{split}
\label{eq:plasmon_cherenkov}
\end{equation}
Here, the geometry (Fig. \ref{fig:plasmon_CR}a) is parameterized as $q' = \sqrt{q_\perp'^2 + q_\parallel'^2}$, where $q_\perp' = -q_\perp$, $q_\parallel' = (\omega + \omega')/\beta c - q_\parallel$, and $(q_\parallel, q_\perp) = (q\cos\theta, q\sin\theta)$. Additionally, $g(q) = (1-e^{-qd})/q$ is a geometric factor which comes from the finite thickness of the film in the $z$-direction. The frequency scale is set by $\omega_0 \equiv \sqrt{2}v/\sigma$, using the same definition as for the free-space Cherenkov emission. Finally, $\Im r_p(\omega, q)$ is the imaginary part of the p-polarized reflection coefficient for the interface above the material \cite{novotny2012principles}. This expression can be easily modified to account for a different distribution of index perturbation in the $z$-direction. The result of Eq. \ref{eq:plasmon_cherenkov} can be separated into interpretable pieces. The exponential factor sets $\omega_0$ as the scale of frequencies which can be excited by the pulse, similarly to the case of two-photon Cherenkov radiation in a homogeneous medium. The factor $(\hat{q}\cdot\hat{q}' - 1)^2$ is responsible for controlling the angular relation between the two plasmons. The imaginary part of the reflectivity peaks where modes are present. There are two factors of $\Im r_p$ --- one for each plasmon --- so that the total expression is peaked for values of $(\omega, q)$ and $(\omega', q')$ which both lie along the dispersion. In the lossless limit, $\Im r_p(\omega, q) \propto q\delta\left(q - q(\omega)\right)$, so both points lie exactly on the dispersion $q(\omega)$. With loss present, this condition is relaxed so that the two frequencies and wavevectors instead lie near the dispersion, set by its linewidth. Models for the optical properties of graphene are taken from \cite{koppens2011graphene, jablan2009plasmonics, christensen2017classical}.

Similarly to a homogeneous medium, phase matching and momentum conservation place constraints on the emitted angles and frequencies of the two plasmons. Fig. \ref{fig:plasmon_CR}b shows the kinematic constraints between the two emitted plasmons assuming a simplified lossless Drude model of the surface conductivity (i.e. assuming that the plasmon has a dispersion $\omega \propto \sqrt{q}$). The two axes show the angle of emission of each of the two plasmons measured with respect to the direction of propagation. If the first plasmon is emitted at an angle which satisfies, $0 \leq \theta \leq \pi$, then the second must satisfy $\pi \leq \theta' \leq 2\pi$ in order to obey total momentum conservation in the direction transverse to propagation. The contours show the kinematic constraints for different values of the dimensionless constant $\hbar\omega\beta/(4\alpha E_{\text{F}})$. With higher frequency and velocity, or lower Fermi level, the contours are pushed toward the line $\theta' = \theta + \pi$. Additionally, the colorbar shows the ratio of plasmon frequencies $\omega/\omega'$, where $\omega$ and $\omega'$ are respectively the lower and higher of the frequencies, indicating the frequency-angle entanglement between the two plasmons. 

Figure \ref{fig:plasmon_CR}c shows the width-normalized angular spectrum $(1/W) d\Gamma/d\omega$ for 2-plasmon emission for two different velocities $\beta = \{0.4, 0.8\}$. We have computed these results for three different optical conductivity models for graphene (Drude, local relaxation time approximation, and nonlocal) in order to gain insights as to how dissipation, dispersion, intraband transitions, and nonlocality impact the production of plasmon pairs. Different doping levels $E_{\text{F}}$ for the graphene sample are also considered. For the more highly doped samples considered (i.e. $E_{\text{F}} = 0.5, 0.8$ eV), the Drude conductivity model is seen to provide a good approximation to the more physically accurate conductivity models. However, intraband transitions and nonlocal effects become important in samples with less doping. Notably, for sufficiently high velocities ($\beta = 0.8$), and relatively low doping levels ($E_{\text{F}} = 0.2$ eV), nonlocal corrections to the conductivity become important. In this regime, particle-hole pairs in the region $\hbar\omega > 2E_{\text{F}}$ are produced. We now comment on the total rate of this process, making a comparison to the free-space case discussed in the main text. To compare the 3D and 2D processes on equal footing, we consider the pair production rate \emph{per unit volume} of dielectric which has its permittivity changed. Accounting for the substrate height and pulse width $\sigma$ in both cases, we have $\Gamma/V \approx 10^4\,$ s$^{-1}\,\mu$m$^{-3}$ for the homogeneous medium, and $\Gamma/V \approx 2\times 10^8\,$ s$^{-1}\,\mu$m$^{-3}$ for plasmons. This corresponds to a four order of magnitude increase in efficiency, showing that 2D polaritonic platforms such as graphene may hold promising opportunities for the observation and utilization of this two-photon process. The ability to electrically tune the dispersion to change the angular and frequency correlations of emission is also an attractive feature of graphene-based systems. 

\section{Supplementary concept figure}

\begin{figure}[h]
    \centering
    \includegraphics{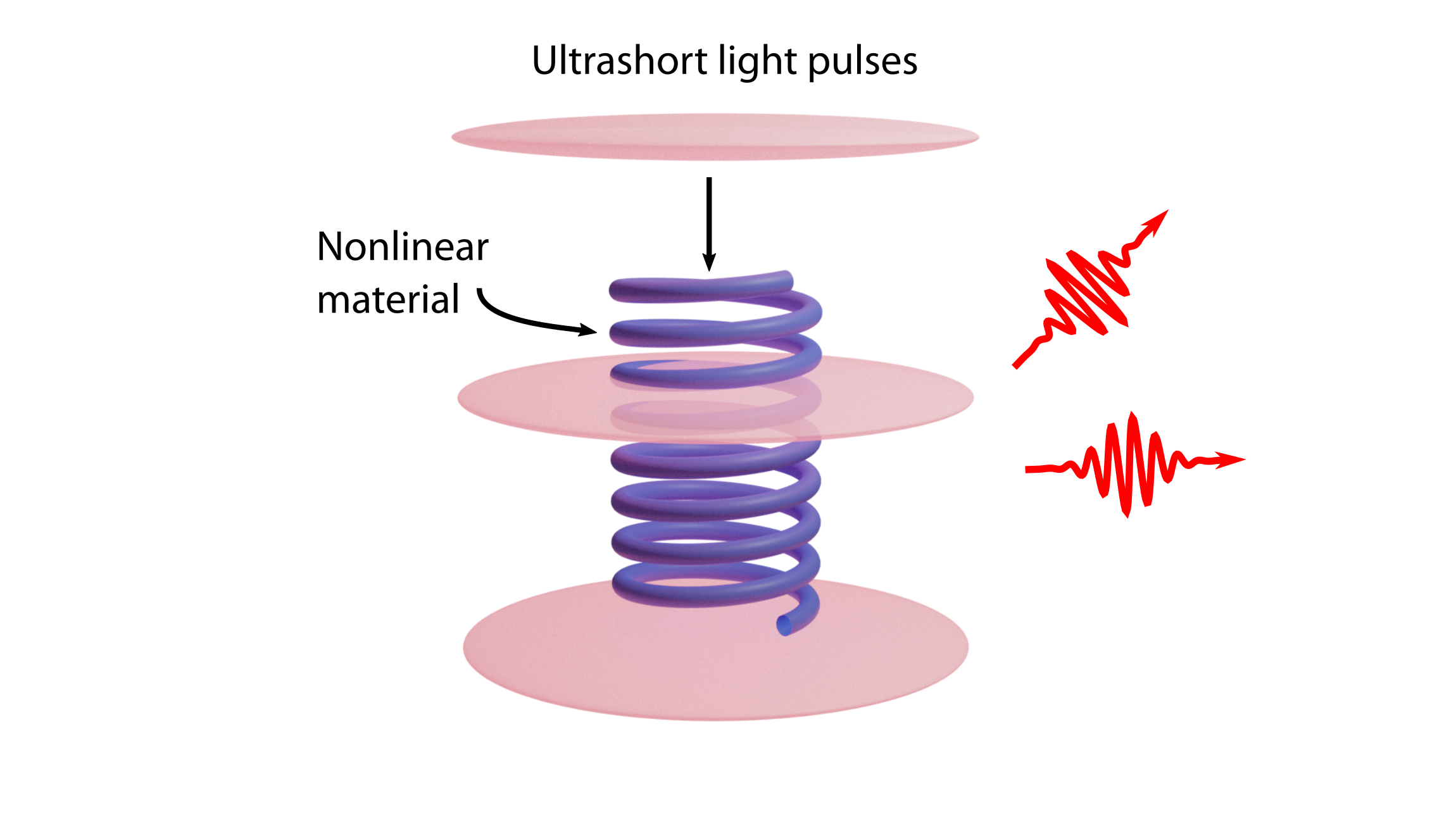}
    \caption{\textbf{Concept figure for for colliding ultrashort light pulses with nonlinear structures to create accelerating index perturbations.} In this particular example, we show a spring-shaped structure which is made of a nonlinear material. When a short pulse collides with the structure, the intersection creates an index perturbation which follows the helical trajectory defined by the structure. Such an index perturbation can travel superluminally, and realize acceleration with a periodicity as described in the ``synchrotron'' geometry (main text Fig. 3), thus emitted entangled photon pairs.}
    \label{fig:slinky}
\end{figure}

\section{Derivations for example systems}
\label{sec:derivations}

\subsection{Derivation of quantum Cherenkov radiation in a homogeneous medium}

Here, we provide a derivation of main text Eq. 4 using the mode expansion method outlined in the first section. We use the index perturbation
\begin{equation}
    \Delta\varepsilon(\rv, t) = \delta\varepsilon\,\exp\left[-\frac{(z - v_0 t)^2}{2\sigma^2}\right].
\end{equation} 
To compute the emission rate, we compute the Fourier transform associated with this index change, defined as $\Delta\varepsilon(z,\omega) = \int_{-\infty}^\infty dt\, e^{i\omega t}\Delta\varepsilon(z, t)$. This is a Gaussian integral, given as 
\begin{equation}
    \Delta\varepsilon(z, \omega) = \delta\varepsilon\,e^{-z^2/2\sigma^2}\sqrt{\frac{\pi}{(v_0^2/2\sigma^2)}}\exp\left[\frac{(i\omega + v_0 z/\sigma^2)^2}{4(v_0^2/2\sigma^2)}\right] = \delta\varepsilon \sqrt{\frac{4\pi}{\omega_0^2}}e^{-\omega^2/\omega_0^2}e^{i(\omega/v_0)z},
    \label{eq:moving_wall_FT}
\end{equation}
where in the second step, we have simplified, and defined $\omega_0^2 \equiv 2v_0^2/\sigma^2$. We see that in frequency space, this perturbation is a plane wave, much like the current distribution of a free electron which produces Cherenkov radiation. In a homogeneous medium of constant nondispersive background index $n = \sqrt{\varepsilon}$, the eigenmodes are expressed as 
\begin{equation}
    \mathbf{F}_k(\rv) = \frac{\hat{\epsilon}_k}{\sqrt{V n^2}} e^{i\kv\cdot\rv},
\end{equation}
where $\hat{\epsilon}_k$ is a polarization, and $V$ is the mode normalization volume . These modes obey the dispersion relation $\omega_k = ck/n$. Then, following Eq. \ref{eq:s_matrix}, the $S$-matrix element which connects the initial ground state $\ket{i} = \ket{0}$ to a 2-photon final state $\ket{f} = \ket{k,k'}$ is
\begin{equation}
\begin{split}
    S_{fi} = \frac{2}{V}\left(\frac{\delta\varepsilon}{\varepsilon}\right)(2\pi)^3\frac{\sqrt{\pi\omega_k \omega_{k'}}}{\omega_0} &\exp\left[-\frac{(\omega_k + \omega_{k'})^2}{\omega_0^2}\right] \left(\hat{\epsilon}_{k}^* \cdot \hat{\epsilon}_{k'}^* \right) \\
    &\times\delta^{(2)}(\kv_\perp + \kv_\perp')\delta\left(\frac{\omega_k + \omega_{k'}}{v_0} - (k_z + k_z')\right).
\end{split}
\end{equation}
Here, the delta functions make momentum conservation and phase matching manifest. The probability of the process is given by $P = \frac{1}{2}\sum_{k,k'}|S_{fi}|^2$, where the factor of $1/2$ accounts for the indistinguishability of the two photons. Defining the transition rate as the probability per unit time, we can write the transition rate $\Gamma$ per unit area $A$ of the wall as
\begin{equation}
    \frac{\Gamma}{A} = \frac{\pi}{2}\left(\frac{\delta\varepsilon}{\varepsilon}\right)^2 \frac{(2\pi)^3 v_0}{\omega_0^2V^2}\sum_{k,k'}|\hat{\epsilon}\cdot\hat{\epsilon}'|^2 \omega_k\omega_{k'} e^{-\frac{2(\omega_k + \omega_{k'})^2}{\omega_0^2}} \delta^{(2)}(\mathbf{k}_\perp + \mathbf{k}_\perp')\delta\left(\frac{\omega_k + \omega_{k'}}{v_0} - (k_z + k_z')\right),
\end{equation}
where we have used $\delta(\omega=0) = T/2\pi$ to define the time of interaction, and $\delta(\kv_\perp=0) = A/(2\pi)^2$ to define the area of the semi-infinite moving wall. We also use the shorthand $\hat{\epsilon}$ to denote $\hat{\epsilon}_\k$, and likewise for the primed version. The polarization factor in the rate indicates that the photons prefer to be oriented in the same direction, and cannot be emitted orthogonally. The mode sums are converted to a continuum as 
\begin{equation}
    \sum_{k,k'} \to \frac{1}{V^2} \sum_{\epsilon,\epsilon'} \int \frac{d^3k}{(2\pi)^3} \frac{d^3k'}{(2\pi)^3}.
\end{equation}
Then, the polarization sum is handled as
\begin{equation}
    \sum_{\epsilon,\epsilon'} |\hat{\epsilon}\cdot\hat{\epsilon}'|^2 = 1 + (\hat{k}\cdot\hat{k}')^2.
\end{equation}
For the transverse delta function, we can write
\begin{equation}
    \delta^{(2)}(\mathbf{k}_\perp + \mathbf{k}_\perp') = \frac{1}{k_\perp}\delta(k_\perp - k_\perp')\delta(\phi - \phi' - \pi).
\end{equation}
We also have $k_z = k\cos\theta$ and likewise for the primed version. This allows us to write
\begin{equation}
\begin{split}
     \frac{\Gamma}{A} = v_0 \frac{\pi}{2}\left(\frac{\delta\varepsilon}{\varepsilon}\right)^2 &\frac{1}{\omega_0^2(2\pi)^2} \int_0^\infty dk\,k^2\,dk'\,k'^2 \int_{-1}^1 d(\cos\theta)\,d(\cos\theta') e^{-\frac{2(\omega_k + \omega_{k'})^2}{\omega_0^2}} \omega_k \omega_{k'}\\
    &\times\left(1 + (\hat{k}\cdot\hat{k}')^2\right)\frac{\delta(k\sin\theta-k'\sin\theta')}{k\sin\theta} \delta\left(\frac{\omega_k + \omega_{k'}}{v_0} - (k\cos\theta + k'\cos\theta')\right)
\end{split}
\end{equation}
The two delta-function conditions must be satisfied simultaneously. These equations (after using the dispersion relation $\omega = ck/n$ to convert $k$'s to $\omega$'s) are
\begin{align}
    \frac{\omega + \omega'}{n\beta} &= \omega\cos\theta + \omega'\cos\theta \label{eq:phase_match} \\ 
    \omega\sin\theta &= \omega'\sin\theta' \label{eq:momentum_conserv}
\end{align}
There is a trivial solution which corresponds to the onset of the radiation. This occurs when $\cos\theta = \cos\theta' = 1/n\beta$. Consequently, $\omega = \omega'$ here. Thus this solution corresponds to 2 photons of the same frequency produced at the classical Cherenkov angle. The pair of equations, however, is singular in $\omega, \omega'$. Taking the determinant of the system and setting to zero reveals an underlying constraint on the angles, which is 
\begin{equation}
    \sin\theta + \sin\theta' = n\beta\sin(\theta + \theta')
\end{equation}
Performing the delta function integrals, we find a final differential decay rate per unit frequency and angle (and normalized to the area $A$) as
\begin{equation}
    \frac{1}{A}\frac{d\Gamma}{d\omega d\theta} = \frac{v_0}{8\pi\omega_0^2}\left(\frac{\delta\varepsilon}{\varepsilon}\right)^2\left(\frac{n}{c}\right)^3 (\omega\omega')^2 e^{-2(\omega+\omega')^2/\omega_0^2} \frac{\left(1 + (\cos\theta\cos\theta' - 1)^2\right)}{\left(\frac{\cot\theta}{n\beta} - \frac{1}{\sin\theta}\right)},
\end{equation}
which is the result quoted in main text Eq. 4.

\subsection{Derivation of quantum Cherenkov radiation for plasmons}
\label{subsec:plasmon_derivation}

Here, we provide details of the derivation of Eq. \ref{eq:plasmon_cherenkov}. We consider an index perturbation 
\begin{equation}
    \Delta\varepsilon(x, t) = \delta\varepsilon\,\exp\left[-\frac{(x - v_0 t)^2}{2\sigma^2}\right],
\end{equation}
for $0 < z < d$, and zero otherwise. The Fourier transform follows a form identical to that of  Eq. \ref{eq:moving_wall_FT}. Then, implementing Eq. \ref{eq:master_prob} with the form of the Green's function shown in Eq. \ref{eq:ImG}, we obtain
\begin{equation}
\begin{split}
    \frac{dP}{d\omega d\omega'} = \frac{\delta\varepsilon^2}{\omega_0^2} e^{-2(\omega+\omega')^2/\omega_0^2}(2\pi) WL &\int\frac{d^2q}{(2\pi)^2}\frac{d^2q'}{(2\pi)^2}\left(\frac{1 - e^{-(q+q')d}}{q + q'}\right)^2 qq' \Im r_p(\omega,q) \Im r_p(\omega',q') \\
    &\times \tr\left[\hat{\epsilon}_{q'}^*\otimes \hat{\epsilon}_{q'} \cdot \hat{\epsilon}_{q} \otimes \hat{\epsilon}_{q}^*\right]\delta(q_\perp + q_\perp')\delta\left(\frac{\omega_q + \omega_{q'}}{v_0} - (q_\parallel + q_\parallel')\right).
\end{split}
\end{equation}
here, $L$ is a length factor which comes from the delta function in the parallel direction. It is related to the time factor $T$ by $L = \beta cT$. Performing the integral $\int d^2q' = \int dq_\parallel'\,dq_\perp'$, and using
\begin{equation}
    \tr\left[\hat{\epsilon}_{q'}^*\otimes \hat{\epsilon}_{q'} \cdot \hat{\epsilon}_{q} \otimes \hat{\epsilon}_{q}^*\right] = \frac{1}{4}\left(\hat{q}\cdot\hat{q}' - 1\right)^2
\end{equation}
we obtain the general result
\begin{equation}
\begin{split}
    \frac{1}{W}\frac{d\Gamma}{d\omega d\theta} = \frac{\beta c}{(2\pi)^3} \frac{\delta\varepsilon^2}{\omega_0^2}\int d\omega' \int dq\,q^2 q'\, e^{-2(\omega+\omega')^2/\omega_0^2}&\left(\frac{1 - e^{-(q+q')d}}{q + q'}\right)^2 (\hat{q}\cdot\hat{q}' - 1)^2 \\
    &\times\Im r_p(\omega, q) \Im r_p(\omega', q'),
\end{split}
\end{equation}
where we have $q' = \sqrt{q_\perp'^2 + q_\parallel'^2}$, where $q_\perp' = -q_\perp$, $q_\parallel' = (\omega + \omega')/\beta c - q_\parallel$, and $(q_\parallel, q_\perp) = (q\cos\theta, q\sin\theta)$. The relation between $q'$ and $q$ in terms of emission frequencies and angles is thus given by
\begin{equation}
    q'^2 = q^2 \sin^2\theta + \left(\frac{\omega+\omega'}{\beta c} - q\cos\theta\right)^2.
\end{equation}
For numerical implementation, we also note that the overlap between wave vectors can be expressed in terms of components as 
\begin{equation}
    \hat{q}\cdot\hat{q}' = \frac{q_\parallel q_\parallel' + q_\perp q_\perp'}{q q'}.
\end{equation}

\subsection{Derivation of two-photon Synchrotron radiation}

Here, we provide details of the derivation of Eq. 6 of the main text, for two-photon synchrotron radiation. For an index perturbation which follows a circular trajectory, we use
\begin{equation}
    \Delta\varepsilon(\rv, t) = \delta\varepsilon \,e^{-\frac{1}{2\sigma^2}(\rho^2 + z^2 + R^2)}e^{\frac{R\rho}{\sigma^2}\cos(\theta - \Omega t)},
\end{equation}
where $(\rho, \theta, z)$ are cylindrical coordinates, $\Omega$ is the angular frequency of precession, $\sigma$ is the Gaussian width, and $R$ is the radius of the circle. Throughout this derivation, we will make use of the Jacobi-Anger expansion, shown in two forms here:
\begin{align}
    e^{iz\cos\theta} &= \sum_{n=-\infty}^\infty e^{in\theta} i^n J_n(z) \\
    e^{z\cos\theta}  &= \sum_{n=-\infty}^\infty e^{in\theta} I_n(z)
\end{align}
Here, $I_n(x) = i^{-n} J_n(ix)$ is a modified Bessel function of the first kind, in terms of the Bessel function of the first kind $J_n(x)$. Using the second form, we can take the Fourier transform of the perturbation as
\begin{equation}
    \Delta\varepsilon(\rv, \omega) = \delta\varepsilon \,e^{-\frac{1}{2\sigma^2}(\rho^2 + z^2 + R^2)} \sum_{n=-\infty}^\infty e^{-in\theta}I_n\left(\frac{R\rho}{\sigma^2}\right) 2\pi\delta(\omega - n\Omega),
\end{equation}
Then, to compute the scattering matrix element, we have that
\begin{equation}
\begin{split}
        \int d^3r\,\Delta\varepsilon(\rv, \omega_{k} + \omega_{k'}) \mathbf{F}_k^*(\rv) \cdot \mathbf{F}_{k'}^*(\rv) = \frac{\delta\varepsilon}{\varepsilon V}(\hat{\epsilon}_k^* \cdot \hat{\epsilon}_{k'}^*) & (2\pi)^{5/2} \sigma^3 e^{-(\kv + \kv')^2\sigma^2/2} \\
        &\times\sum_{m=-\infty}^\infty J_m(KR) \delta(\omega_k + \omega_{k'} - m\Omega).
\end{split}
\end{equation}
Here, $K = \sqrt{(k_x + k_x')^2 + (k_y + k_y'^2)^2}$ is the net in-plane wavevector of emission. As an intermediate step, we have made use of the integral
\begin{equation}
    \int_0^\infty d\rho\, \rho J_m(K\rho)I_m\left(\frac{R\rho}{\sigma^2}\right) e^{-\frac{\rho^2}{2\sigma^2}} = \sigma^2 e^{-\frac{\sigma^2K^2}{2}} e^{\frac{R^2}{2\sigma^2}} J_m(KR)
\end{equation}
By squaring the matrix element, and summing over final two-photon states, the probability of two-photon emission from this system per unit time $T$ can be written as
\begin{equation}
    \begin{split}
        \frac{P}{T} = \left(\frac{\delta\epsilon}{\epsilon}\right)\frac{\sigma^6}{32\pi^2} \sum_{m=-\infty}^\infty \int \frac{d^3k}{(2\pi)^3} \frac{d^3k'}{(2\pi)^3} &\omega_k \omega_{k'} \left(1 + (\hat{\kv}\cdot\hat{\kv}')^2\right) \\
        &\times e^{-\sigma^2\left(\kv+\kv'\right)^2} J_m^2(KR) \delta(\omega_k + \omega_{k'} - m\Omega)
    \end{split}
\end{equation}
After some algebra, we find that the differential rate of down-conversion per unit angle $\theta$ and frequency $\omega$ into each harmonic $m$ is given as:
\begin{equation}
       \frac{d\Gamma_m}{d\omega d\theta} = \left(\frac{\delta\varepsilon}{\varepsilon}\right)^2 \frac{\sigma^6 n^6 \omega^3 (m\Omega - \omega)^3\sin\theta}{16\pi c^6} \int d\Omega' \, \left(1 + (\hat{\kv}\cdot\hat{\kv}')^2\right)e^{-\sigma^2\left(\kv+\kv'\right)^2} J_m^2(KR).
\end{equation}
The total differential rates can be found by summing over many harmonics $m$. 

\section{Green's functions in layered media}
\label{sec:green_function}

In the derivation of Cherenkov-like radiation into graphene plasmons, we used expressions for Green's functions for polaritons on layered surfaces. We provide a summary of this method here. In a dielectric with permittivity $\varepsilon(\rv, \omega)$, the Maxwell dyadic Green's function $G(\rv, \rv', \omega)$ is defined so that
\begin{equation}
    \left(\curl\curl - \frac{\omega^2}{c^2}\varepsilon(\rv, \omega)\right)G(\rv, \rv', \omega) = \delta^{(3)}(\rv - \rv') I,
    \label{eq:DGF}
\end{equation}
where $I$ is the $3\times 3$ identity matrix. In a 2D translation-invariant medium (such as in a geometry of planar slabs), the Green's function can be written as a Fourier transform in the invariant directions (denoted by $\rho$) as 
\begin{equation}
    G(\rv, \rv', \omega) = \int \frac{d^2q}{(2\pi)^2} e^{i\qv\cdot(\rho-\rho')} g_{\qv}(z,z',\omega),
\end{equation}
where $g_\qv(z,z',\omega)$ is now a reduced Green's function for the $z$-direction only. Substituting this expansion into Eq. \ref{eq:DGF}, we find that the reduced Green's function satisfies the equation
\begin{equation}
    \left((\partial_z \hat{z} + i\qv)\times(\partial_z \hat{z} + i\qv)\times - \frac{\omega^2}{c^2}\varepsilon(z, \omega)\right)g_\qv(z,z',\omega) = \delta(z-z')I
    \label{eq:DGF_reduced_eq}
\end{equation}
If the final interface is at $z=0$, and we have evanescently decaying waves in the electrostatic limit, for points $\rv, \rv'$ above the interface, we have the solution
\begin{equation}
    g_\qv(z,z',\omega) = e^{-q(z+z')} \frac{c^2 q}{\omega^2} r_p(\omega, q) \hat{\epsilon}_q\otimes \hat{\epsilon}_q^*,
    \label{eq:p_pol_g}
\end{equation}
where $\hat{\epsilon}_q = (\hat{q} + i\hat{z})/\sqrt{2}$, and $r_p(\omega, q)$ is the $p$-polarized reflectivity of the surface \cite{novotny2012principles}. 

We can now use this expression, along with properties of $g_\qv$, to obtain a useful form for $\Im G(\rv,\rv',\omega')$. From the reciprocity of $G$, it follows that $g_{-\qv}(z,z',\omega) = g_\qv^T(z',z,\omega)$. This allows us to write
\begin{align}
    \Im G(\rv, \rv', \omega) &= \Im \int \frac{d^2q}{(2\pi)^2} e^{i\qv\cdot(\rho-\rho')} g_{\qv}(z,z',\omega) \\
    &= \int \frac{d^2q}{(2\pi)^2} e^{i\qv\cdot(\rho-\rho')}\left(\frac{ g_{\qv}(z,z',\omega) - g_{\qv}^\dagger(z',z,\omega)}{2i}\right).
\end{align}
Taking the dagger of $g_\qv$ as defined in Eq. \ref{eq:p_pol_g} leaves the polarization factor unchanged, and results in conjugation only on the reflection coefficient. This allows us to write the form
\begin{equation}
    \Im G(\rv, \rv', \omega) = \int\frac{d^2q}{(2\pi)^2} e^{i\qv\cdot(\rho-\rho')}e^{-q(z+z')} \frac{c^2 q}{\omega^2} \Im r_p(\omega, q) \hat{\epsilon}_q\otimes \hat{\epsilon}_q^*,
    \label{eq:ImG}
\end{equation}
which is used to derive Eq. \ref{eq:plasmon_cherenkov}

\bibliography{vacuum.bib}